\documentclass[amssymb, nofootinbib, superscriptaddress]{revtex4}

\usepackage{amsmath}
\usepackage{amsfonts}
\usepackage{graphicx}
\usepackage{psfrag}
\usepackage{array}
\usepackage{bbm}
\newcommand{\f}{\frac}
\newcommand{\tr}{\mathrm{tr}}

\newcommand{\su}{\mathfrak{su}}
\newcommand{\an}{\mathfrak{an}}
\newcommand{\so}{\mathfrak{so}}
\newcommand{\SU}{\mathrm{SU}}
\newcommand{\AN}{\mathrm{AN}}

\newcommand{\ISU}{\mathrm{ISU}}
\newcommand{\SO}{\mathrm{SO}}
\newcommand{\SL}{\mathrm{SL}}

\newcommand{\Spin}{\mathrm{Spin}}

\newcommand{\R}{\mathbb{R}}
\newcommand{\C}{\mathbb{C}}
\newcommand{\N}{\mathbb{N}}
\newcommand{\Z}{\mathbb{Z}}

\newcommand{\cM}{{\cal M}}

\newcommand{\uu}{{\cal U}}
\newcommand{\ttt}{{\cal T}}
\newcommand{\tF}{{\tilde F}}

\newcommand{\id}{\mathbb{I}}
\newcommand{\be}{\begin{equation}}
\newcommand{\ee}{\end{equation}}
\newcommand{\bes}{\begin{eqnarray}}
\newcommand{\ees}{\end{eqnarray}}

\def\cc{{\cal C}}

\def\kk{{\cal K}}
\def\lll{{\cal L}}
\def\mm{{\cal M}}
\def\nn{{\nonumber}}

\def\ss{{\cal S}}

\def\uu{{\cal U}}

\def\mm{{\cal M}}
\def\tl{\widetilde}

\def\arr{\rightarrow}
\def\eps{\epsilon}

\def\ka{\kappa}
\def\cc{{\cal C}}
\def\la{\langle}
\def\ra{\rangle}

\renewcommand{\hat}{\widehat}
\def\kk{{\cal K}}

\def\ss{{\cal S}}
\def\tlF{\tl{F}}
\newcommand{\mat}[2]{\left(\begin{array}{#1} #2\end{array}\right)}
\newcommand{\act}{\triangleright}
\newcommand{\one}{\mbox{$1 \hspace{-1.0mm}  {\bf l}$}}
\def\mn{{\mu\nu}}
\newcommand{\mone}{^{-1}}
\def\cop{\Delta}

\def\x{\mathfrak{x}}

\def\tLambda{\tl{\Lambda}}
\def\dx{{\bf [dx^d]}}
\def\meas{{\bf [d\mu]}}
\def\x{{\bf x}}
\def\mn{{\mu\nu}}
\def\mmm{{\mathfrak{m}}}
\newcommand{\Ref}[1]{(\ref{#1})}
\def\nn{\nonumber}

\begin{document}
%
\title{\large\bf 4d Deformed Special Relativity from Group Field Theories}

\author{Florian Girelli}\email{girelli@physics.usyd.edu.au}
\affiliation{SISSA, Via Beirut 2-4, 34014 Trieste, Italy and INFN, Sezione di Trieste}
\affiliation{School of Physics, The University of Sydney, Sydney, New South Wales 2006, Australia}
\author{Etera R. Livine}\email{etera.livine@ens-lyon.fr}
\affiliation{Laboratoire de Physique, ENS Lyon, CNRS UMR 5672, 46 All\'ee d'Italie, 69007 Lyon, France}

\author{Daniele Oriti}\email{daniele.oriti@aei.mpg.de}
\affiliation{Perimeter Institute for Theoretical Physics, 31 Caroline
St, Waterloo, Ontario N2L 2Y5, Canada}
\affiliation{Institute for Theoretical Physics, Utrecht University, Leuvenlaan 4, Utrecht 3584 TD, The Netherlands, EU}
\affiliation{Albert Einstein Institute, Am Muehlenberg 4, Golm,
Germany, EU}

\date{\small \today}


\begin{abstract}
We derive a scalar field theory of the deformed special relativity type, living on non-commutative $\ka$-Minkowski spacetime and with a $\ka$-deformed Poincar\'e symmetry, from the $\SO(4,1)$ group field theory defining the transition amplitudes for topological BF-theory in 4 space-time dimensions. This is done at a non-perturbative level of the spin foam formalism working directly with the group field theory (GFT). We show that matter fields emerge from the fundamental model as perturbations around a specific phase of the GFT, corresponding to a solution of the fundamental equations of motion, and that the non-commutative field theory governs their effective dynamics.

\end{abstract}

\maketitle

\section{Introduction}
\label{intro}
The progress toward a quantum theory of gravity, in the past twenty years or
so, has been substantial. On the theory side, many different approaches, the most
notable being probably string theory, have been developed and
achieved considerable successes \cite{libro}.

Group field theories \cite{gftdaniele,gftlaurent} are quantum field theories over group manifolds, characterized by a non-local pairing of field arguments in the action, which can be seen as a generalization of matrix models \cite{mam}. The combinatorics of the field arguments in
the interaction term of the group field theory (GFT) action follows that of (D-2) faces of a D-simplex, with the GFT field itself interpreted as a (second) quantization of a (D-1)-simplex. The kinetic term of the action governs the  gluing of two D-simplices along a common (D-1)-simplex. Because of this combinatorial structure, the GFT Feynman diagrams, themselves cellular complexes, are dual to D-dimensional simplicial complexes. Thus GFTs can be seen as a simplicial \lq\lq third quantization\rq\rq  of gravity \cite{3rd}, in which a discrete spacetime emerges as a Feynman diagram of the theory in perturbative expansion. The field arguments assign group-theoretic data to these cellular complexes, and the GFT perturbative expansion in Feynman
amplitudes define uniquely and completely  a so-called spin
foam model \cite{gftcarlo}.
Spin foam models \cite{SF}, in turn, can be understood as a covariant formulation
of the dynamics of loop quantum gravity \cite{LQG} and as a new
algebraic implementation of discrete quantum gravity approaches, such as Regge calculus \cite{williams}
and dynamical triangulations \cite{DT}. This makes GFTs a very useful tool, and suggests that they may provide the
fundamental definition of a dynamical theory of spin networks, and be of great help in investigating non-perturbative and collective properties of their
quantum dynamics \cite{gftlaurent, gftdaniele, daniele}.

In recent years, moreover, the possibility of testing
experimentally Planck scale effects using astrophysical or
cosmological observations has been investigated to a great extent
and led to a whole set of approaches to possible quantum gravity
phenomenology \cite{QGPhen}. The general idea is that there exist
several physical amplifying mechanisms, e.g. in gamma-ray bursts,
cosmic rays, or gravitational wave physics, that could bring
quantum gravity effects, even if suppressed by (negative) powers
of the Planck energy or by (positive) powers of the Planck length,
within reach of near future (if not current, e.g. the on-going GLAST
experiment) experiments. The most studied effects are
that of a breaking (e.g. Einstein-Aether theory) or of a
deformation (e.g. Deformed Special Relativity) of fundamental
spacetime symmetries, like the Lorentz or Poincar\'{e} invariance \cite{QGPhen}.
This last case is implemented in the context of non-commutative models of spacetime, with symmetry groups implemented by means of appropriate Hopf algebras \cite{majid1}. In many of the interesting cases, in particular those we are concerned with in this work, spacetime coordinates, turned into operators, have Lie algebra-type commutation relations, with a corresponding momentum space given instead by a group manifold, following the general principle \cite{majid1} that non-commutativity in configuration space is related to curvature in momentum space, a sort of \lq\lq co-gravity\rq\rq  \cite{majid1}. One class of models that have attracted much attention in this context is given by so-called Deformed (or Doubly) Special Relativity \cite{DSR, QGPhen}, based on the idea of introducing a second invariant scale, given by the Planck length (or energy) and assumed to encode quantum gravity effects in a semi-classical and flat spacetime, on top of the velocity scale of usual Special Relativity, while maintaining the relativity principle, and thus a 10-dimensional transformation group relating the observations made by inertial observers. In one particular incarnation of DSR, spacetime is non-commutative and its structure is of $\kappa$-Minkowski type \cite{ruegg}. This is the spacetime we are concerned with here.
In fact, it is now clear that these effective models of quantum gravity can in
principle be falsified. Unfortunately, we are still lacking any fundamental formulation of quantum gravity that, on
top of being clearly defined at the Planck scale,
can produce unambiguously any of the effective models that have
been proposed, thus producing falsifiable predictions.

Very interesting results have been obtained in the 3d context \cite{PR1,PR3,effqg} where it has been shown that effective models with quantum group symmetries and a non-commutative spacetime structure (although different from the DSR one) arise very naturally when considering the coupling of point particles to a spin foam model for 3d quantum gravity, in the Riemannian setting, with the physics of these particles being that of non-commutative field theories on Lie algebra spaces. While no similarly solid links between spin foam models and non-commutative field theories have been discovered in the 4d context, several arguments have been put forward suggesting that these links should exist and that the relevant effective models in 4d should indeed be of the DSR type \cite{heuristic, artem}.

For reasons that should become apparent in the following, group field theories are a natural framework for establishing such links, and for actually {\it deriving} effective non-commutative models of quantum gravity from more fundamental (if tentative) descriptions of quantum spacetime. Once more, in 3d this is technically easier to do, and it has been shown recently \cite{phase} that one can indeed derive the same effective field theory obtained in \cite{PR3} directly from GFT model corresponding to the spin foam model on which that earlier work was based. The procedure used, moreover, appears not to depend too much in the details of the 3d model considered, but only on general properties of the GFT formalism. We will review briefly these results in the next section.

What we do in this paper is to apply the same procedure to the more technically challenging case of four spacetime dimensions, and Lorentzian signature, and derive from a group field theory model related to 4-dimensional quantum gravity an effective non-commutative field theory of the DSR type and living on $\kappa$-Minkowski spacetime.

As said, not only this is the first example of a derivation of a DSR model for matter from a more fundamental quantum gravity model, and one further example of the link between non-commutative geometry and quantum gravity formulated in terms of spin foam/loop quantum gravity ideas, but it is of great interest from the point of view of quantum gravity phenomenology. It is also interesting, more generally, as another possible way of bridging the gap between quantum gravity at the Planck scale and effective physics at low energies and macroscopic distances.

\section{A brief review of 3d group field theory and effective theories}
The group field theory generating the Ponzano-Regge spinfoam
amplitudes for 3d quantum gravity was written by Boulatov
\cite{boulatov}. The field $\phi: \SU(2)^3\rightarrow \C$  is
required to be gauge invariant under the diagonal right action of
$\SU(2)$: \be \phi(g_1,g_2,g_3)=\phi(g_1g,g_2g,g_3g), \quad\forall
g\in\SU(2). \ee The action defining this 3d group field theory
involves a trivial propagator and the tetrahedral interaction
vertex: \be S_{3d}[\phi]\,=\, \f12\int[dg]^3
\phi(g_1,g_2,g_3)\phi(g_3,g_2,g_1) -\f{\lambda}{4!}\int [dg]^6
\phi(g_1,g_2,g_3)\phi(g_3,g_4,g_5)\phi(g_5,g_2,g_6)\phi(g_6,g_4,g_1).
\ee This field theory generates Feynman diagrams which are
identified to 3d triangulations. Moreover the evaluation of these
Feynman diagrams gives exactly the Ponzano-Regge amplitude
associated to the corresponding triangulation.

We can require the field to satisfy a reality condition,
$\bar{\phi}(g_1,g_2,g_3)=\phi(g_3,g_2,g_1)$, ensuring the
quadratic kinetic term to be real. Further symmetry requirements
under permutations of the field arguments can be imposed. For
example, a symmetry under even permutations combined with complex
conjugation under odd permutations leads to the corresponding
perturbative expansion to involve only orientable complexes (see
e.g. \cite{dfkr,dp-p}). However, no specific condition is needed
for the definition of the model, nor for our procedure and results
to apply, so we do not discuss this issue further.

As it was shown in \cite{phase}, we can identify a specific type
of fluctuations of the group field $\phi$ as matter degrees of
freedom propagating on some effective flat non-commutative
background. More precisely, this effective dynamics is given by a
non-commutative quantum field theory invariant under a quantum
deformation of the Poincar\'e group. The procedure is simple: we
look at some two-dimensional variations of the $\phi$-field around
classical solutions (as we work in a time-less setting, we could
call these solutions \lq\lq instantons\rq\rq) of the group field
theory.

The equation of motion of the group field theory for this model
are given by: \be \phi(g_3,g_2,g_1)\,=\,\f{\lambda}{3!}\int
dg_4dg_5dg_6 \phi(g_3,g_4,g_5)\phi(g_5,g_2,g_6)\phi(g_6,g_4,g_1).
\ee Calling $\phi^{(0)}$ a generic solution to this equation, we
look at field variations
$\delta\phi(g_1,g_2,g_3)\equiv\psi(g_1g_3^{-1})$ which do not
depend on the group element $g_2$ and which we call
``two-dimensional variations". This leads to an effective action
describing the dynamics of the 2d variations $\psi$ around the
background solution $\phi^{(0)}$: \be
S_{eff}[\psi]\,\equiv\,S_{3d}[\phi^{(0)}+\psi]-S_{3d}[\phi^{(0)}].
\ee This effective action will not contain any (infinite, in
general) constant term (which we cancel by means of the
counter-term $S_{3d}[\phi^{(0)}]$), nor linear terms since
$\phi^{(0)}$ is a classical solution. The first term will be a
non-trivial quadratic, i.e. kinetic, term for the variation
$\psi$, as we will see below.

We consider a specific class of classical solutions, named ``flat"
solutions: \be
\phi^{(0)}(g_1,g_2,g_3)\,=\,\sqrt{\f{3!}{\lambda}}\,\int dg\;
\delta(g_1g)F(g_2g)\delta(g_3g), \quad F:G\rightarrow\R. \ee As
shown in \cite{phase}, this ansatz gives solutions to the field
equations as soon as $\int F^2=1$ (or $F=0$). Also, this type of
solutions can be considered as a \lq\lq regularized or
smoothed\rq\rq version of a function
$\phi^{(0)}(g_1,g_2,g_3)\,=\,\sqrt{\f{3!}{\lambda}}\,\int dg\;
\delta(g_1g)\delta(g_2g)\delta(g_3g) $, formally still a solution
of the field equations (but giving divergent terms when inserted
in the equations) and representing a quantum space in which all
possible $SU(2)$ holonomies are flat, i.e. a \lq\lq quantum flat
space\rq\rq . There exists other solutions \cite{withflo} but they
are not relevant to the present discussion. It is then
straightforward to compute the effective action: \be
S_{eff}[\psi]\,=\,
\f12\int\psi(g)\kk(g)\psi(g^{-1})-\f\mu{3!}\int[dg]^3\,
\psi(g_1)\psi(g_2)\psi(g_3)\delta(g_1g_2g_3)
-\f{\lambda}{4!}\int[dg]^4\, \psi(g_1)..\psi(g_4)\delta(g_1..g_4),
\ee with the kinetic term and the 3-valent coupling given in term
of $F$:
$$
\kk(g)\,=\,1-2\left(\int F\right)^2-\int dh F(h)F(hg),
\qquad
\f\mu{3!}\,=\,\sqrt{\f{\lambda}{3!}}\,\int F.
$$
As it was shown in \cite{PR3,effqg,majid,phase,karim,matrix}, such
an action defines a non-commutative quantum field theory invariant
under a quantum deformation of the Poincar\'e group $\ISU(2)$.

First of all, in order to understand this correspondence, one
should interpret the above action as written in momentum space:
$\SU(2)\sim\ss^3$ is a curved momentum space with a group
structure and the constraints $\delta(g_1g_2g_3)$ and
$\delta(g_1..g_4)$ express momentum conservation in the
interaction. This also implies that the effective theory itself is
a group field theory. The explicit duality between this group
field theory and the non-commutative field theory, i.e. the same
field theory written in terms of a non-commutative configuration
space, is then achieved through a non-commutative Fourier
transform mapping functions on $\SU(2)$ to functions on $\R^3$
endowed with a non-commutative $\star$-product, on $\R^3$, dual to
the convolution product on $\SU(2)$. Following this Fourier
transform, the $\SU(2)$ momentum conservation maps to a deformed
conservation law on $\R^3$ and the kinetic term $\kk(g)$ becomes
the differential operator defining the field's dynamics in terms
of derivatives on $\mathbb{R}^3$ (the Laplacian in the simplest
case). The details of this mapping do not concern us here, as we
are more interested in the general procedure and in how the specific
kinetic terms look like, but it will be described in some detail
in the following sections for the 4-dimensional Lorentzian case,
which is of direct relevance for DSR.

In order to get an action invariant 3d rotations, $\psi(g)\arr\psi(hgh^{-1})$, we usually assume
that the function $F(g)$ is invariant under conjugation $F(g)=F(hgh^{-1})$. Such functions are
linear combination of the $\SU(2)$ characters:
\be
F(g)=\sum_{j\in\N/2} F_j\chi_j(g), \qquad F_0=\int F, \qquad F_j=\int dg\, F(g)\chi_j(g),
\ee
where the $F_j$'s are the Fourier coefficients of the Peter-Weyl decomposition on the field. The
(spin) $j\in\N/2$ label the irreducible representations of $\SU(2)$, which have dimension
$d_j\equiv (2j+1)$. The normalization constraints now reads:
$$
\int F^2 \,=\,1\,=\,
\sum_j F_j^2.
$$
The kinetic term is easily written in term of the $F_j$'s and the characters $\chi_j(g)$:
\be
\kk(g)=1-3F_0^2-\sum_{j\ge 0}\f{F_j^2 }{d_j} \chi_j(g)=\sum_{j\ge 0} F_j^2\left(1-\f{\chi_j(g)}{d_j}\right)-2F_0^2\,\equiv\,
Q^2(g)-M^2.
\ee
It is easy to check that $Q^2(g)\ge 0$ is always positive (since $|\chi_j(g)|\le d_j$) and vanishes
at the identity $g=\id$. We interpret this term as the ``Laplacian" of the theory while the 0-mode
$F_0$ defines the mass $M^2\equiv 2 F_0^2$.

The standard choice is given by the 3-dimensional representation labeled by $j=1$. We parameterize
$\SU(2)$ group elements as two by two matrices:
\be
g=\cos\theta+i\sin\theta\hat{u}\cdot\vec{\sigma},
\ee
where $\theta\in[0,2\pi]$ is the class angle (half of the rotation angle), $\hat{u}\in\ss^2$ is the
rotation axis and the $\vec{\sigma}$'s are the (Hermitian) Pauli matrices. The 3d momentum is
usually defined as the projection of the group element on the Lie algebra $\su(2)$:
\be
\vec{p}\,=\,\f{1}{2i}\tr g\vec{\sigma}\,=\, \sin\theta\hat{u},
\ee
so that the Laplacian would be given by a term $p^2=\sin^2\theta$. Now the characters are functions
of the class angle:
\be
\chi_j(g)=\f{\sin d_j\theta}{\sin\theta}.
\ee
So that the spin-1 case allows to recover the $p^2$ kinetic term:
$$
\chi_1(g)=\f{\sin3\theta}{\sin\theta}=1+2\cos2\theta=3-4p^2.
$$
Finally, choosing a classical solution given entirely by the
character $\chi_1(g)$ up to a constant shift, \be
F(g)=a+b\chi_1(g), \qquad \int F= a^2+b^2=1, \ee we obtain a
simple kinetic term with the Laplacian and a mass: \be
\kk(g)=\f43(1-a^2)\,\vec{p}^2 -2a^2. \ee Clearly, other choices
are possible and they may give rice to higher order differential
operators and thus to more complicated kinetic terms.

\medskip

We conclude this review section by a short remark on the
possibility relaxing the classical solution condition. Consider a
background field given the flat ansatz $\phi^{(0)}=\int \delta
F\delta$, but without requiring it to satisfy the normalization
condition $\int F^2=1$.  The effective action for 2d variations
can still be computed in the same way and what we obtain is a
similar action but with a term linear in the $\psi$-field: \bes
S_{eff}[\psi]&=&\sqrt{\f{3!}{\lambda}}\,\psi(\id)\int dg
F(g)\,\left(1-\int dh F^2(h)\right)+
\f12\int[dg]\psi(g)\kk(g)\psi(g) \\&&-\sqrt{\f{\lambda}{3!}}\,\int
dg F(g)\,\int[dg]^3\, \psi(g_1)..\psi(g_3)\delta(g_1..g_3)
-\f{\lambda}{4!}\int[dg]^4\, \psi(g_1)..\psi(g_4)\delta(g_1..g_4).
\nn \ees We notice that the linear term may disappear even in this
case, i.e. even if $\phi_0$ is not a classical solution, if we
simply assume that $\int F=0$ vanishes. We call this class of
fields satisfying this conditions ``partial classical solutions".
Looking at the effective dynamics of the field $\psi$ around such
partial classical solutions, we still get a quadratic kinetic term
with a non-trivial propagator defined by $F$ and a quartic
interaction term, while the cubic interaction term vanishes as
well.

\section{4d Group Field Theory and perturbations} \label{4dcase}
\label{compact}
The 4-dimensional construction proceeds analogously. We show here
the general form of the class of solutions we deal with, the type
of perturbations we study and which lead to emergent matter
fields, and the general form of the effective actions that result
from the expansion. We will see that this part of the
construction, which works for any group $G$, is straightforward.
The real task, which we tackle in the rest of the paper, will be
to identify the specific example(s) of fundamental GFT actions, of
classical solutions and perturbations, whose effective action are
defined on the specific momentum and configuration space
characterizing DSR theories, i.e. the group $AN_3$ and the
$\ka$-Minkowski non-commutative space respectively, and possess the
right kinetic term, i.e. the one characterized by the appropriate
symmetries.

Let us consider a general 4d GFT related to topological BF quantum
field theories, i.e. whose Feynman expansion leads to amplitudes
that can be interpreted as discrete BF path integrals, for a
compact semi-simple gauge group ${\cal G}$. This is given by the
following action: \bes S_{4d}&=&\f12\int
[dg]^4\,\phi(g_1,g_2,g_3,g_4)\phi(g_4,g_3,g_2,g_1) \\&&
-\f{\lambda}{5!} \int [dg]^{10}
\phi(g_1,g_2,g_3,g_4)\phi(g_4,g_5,g_6,g_7)\phi(g_7,g_3,g_8,g_9)\phi(g_9,g_6,g_2,g_{10})\phi(g_{10},g_8,g_5,g_1),
\nn \ees where the field is required to be gauge-invariant,
$\phi(g_1,g_2,g_3,g_4)=\phi(g_1g,g_2g,g_3g,g_4g)$ for all group
elements $g\in{\cal G}$. The relevant groups for 4d quantum
gravity are ${\cal G}=\Spin(4)$ (and $\SO(5)$) in the Riemannian
case and ${\cal G}=\SL(2,\C)$ (and $\SO(4,1)$) in the Lorentzian
case. In this section, we focus on the compact group case. We will
deal with the non-compact group case relevant to Lorentzian
gravity in the next section. It will require proper and careful
regularization to avoid divergencies due to the non-compact nature
of the group.

We generalize the ``flat solution" ansatz of the 3d group field
theory to the four-dimensional case \cite{phase}: \be
\phi^{(0)}(g_i)\,\equiv\, {}^3\sqrt{\f{4!}{\lambda}}\int
dg\,\delta(g_1g)F(g_2g)\tlF(g_3g)\delta(g_4g). \ee It is
straightforward to check that this provides a solution to the
classical equations of motion as soon as $(\int F\tlF)^3=1$. We
let aside for a moment this normalization condition, and we
compute the effective action for two-dimensional variations around
such background configurations for arbitrary functions $F$ and
$\tlF$:
$$
S_{eff}[\psi]\,\equiv\,S_{4d}[\phi^{(0)}+\psi(g_1g_4^{-1})]-S_{4d}[\phi^{(0)}].
$$
We obtain an effective action with a linear term proportional to
$\psi(\id)$, a non-trivial quadratic kinetic term and interaction
vertices of order 3 to 5: \bes S_{eff}[\psi]&=&
{}^3\sqrt{\f{4!}{\lambda}}\psi(\id)\int F \int \tlF
\left[1-\left(\int F\tlF\right)^3\right] +\f12\int
\psi(g)\psi(g^{-1})\kk(g) \nn\\ && -{}^3\sqrt{\f{\lambda}{4!}}\int
F\int \tlF\,\int\psi(g_1)..\psi(g_3)\,\delta(g_1..g_3) \left[\int
F\int\tlF+\int dh F(hg_3)\tlF(h)\right] \\ &&
-\left({}^3\sqrt{\f{\lambda}{4!}}\right)^2\int F\int \tlF\,\int
\psi(g_1)..\psi(g_4)\,\delta(g_1..g_4) -\f\lambda{5!}\int
\psi(g_1)..\psi(g_5)\,\delta(g_1..g_5), \nn \ees with the new
kinetic operator given by: \be \kk(g)\,=\,\left[1-2\left(\int F
\int \tlF\right)^2\int F\tlF -2 \int F\int \tlF \int dh
F(hg)\tlF(h) \int dh F(h)\tlF(hg)\right]. \ee Taking into account
the normalization condition $(\int F\tlF)^3=1$ and thus working
with an exact solution $\phi_0$ of the equations of motion, we see
that the linear term vanishes exactly due to this condition. We
also notice that, if we were to relax this normalization condition
and work with a \lq\lq partial solution requirement\rq\rq as in
the 3d case, the linear term could still be made to vanish and
with the same condition $\int F=0$ (or with $\int \tlF=0$).
However, in this 4d case, this other condition makes also all new
terms (among which the non-trivial kinetic term) vanish. Another
possibility could be to renormalize the coupling constant
$\lambda$ by re-absorbing in it the factors $\int F\int \tlF$, and
then impose the same condition of vanishing integral in some
limiting procedure. The interest and consequences of doing this,
however, are not clear at the present stage.

At any rate, we obtain an effective field theory for the field
$\psi$ defined on two copies of the initial group manifold, but
reduced by means of the symmetry requirement to a function of a
single group element, with a non-trivial quadratic propagator. The
group ${\cal G}$ is now interpreted again as the momentum space
for the quanta corresponding to this field, with the
$\delta(g_1..g_n)$ factors in the action imposing momentum
conservation in the field interactions. And again, after
introducing a suitable Fourier transform, such effective group
field theory appears as the dual of a non-commutative field
theory. This same duality implies that position space field theory
is defined in terms of functions on $\mathbb{R}^d$, with $d$ the
dimension of the group $G$, endowed with a suitable star product
structure, or, equivalently, by elements of the enveloping algebra
for the Lie algebra of the same group $G$, i.e. non-commutative
fields living on a non-commutative spacetime given by the same Lie
algebra. The non-commutativity reflects the curvature of the group
manifold and the non-abelian group multiplication leads to a
deformation of the addition of momenta. We will show how this
works in detail in the next section for the non-compact group
$G=\SO(4,1)$ and for a group field theory more closely related to
4d quantum gravity.

We conclude this section by considering the special case when the
function $\tF$ is fixed to be the $\delta$-distribution while $F$
is kept arbitrary as long as $F(\id)=1$. This ansatz clearly
satisfies the normalization condition $\int F\tF=1$ and thus
provides a solution to the classical field equations. Calling
$c\equiv\int F$, the effective action takes has a simpler
expression: \bes S_{eff}[\psi]&=& \f12\int
\psi(g)\psi(g^{-1})\left[1-2c^2-2cF(g)F(g^{-1})\right]
-c\left({}^3\sqrt{\f{\lambda}{4!}}\right)\,\int\psi(g_1)..\psi(g_3)\,\delta(g_1..g_3)
\left[c+F(g_3)\right] \nn\\ &&
-c\left({}^3\sqrt{\f{\lambda}{4!}}\right)^2\,\int
\psi(g_1)..\psi(g_4)\,\delta(g_1..g_4) -\f\lambda{5!}\int
\psi(g_1)..\psi(g_5)\,\delta(g_1..g_5). \label{compacteffaction}
\ees

\section{Deformed Special Relativity as a Group Field Theory}
The term "Deformed Special Relativity" (DSR) has been used to describe many different theories. We
are here interested in the original construction which described a non-commutative space-time, of
the Lie algebra type ($\ka$-Minkowski) together with some deformed Poincar\'e symmetries. In
particular these latter are consistent with the existence of another universal scale (the Planck
mass/momentum) than the speed of light.

When dealing with such theory, the literature has often emphasized its non-commutative geometry
aspect. Moreover it is also known  since some time  \cite{majid1} that a Fourier transform from a
non-commutative space-time of the Lie algebra type leads to a (curved) momentum space with a
(non-abelian) group structure. From this perspective, it is clear that a scalar field theory over
$\ka$-Minkowski can also be interpreted as a group field theory, where the group is the momentum
space (contrary to the usual GFT approach for quantum gravity models where the group is usually
considered as the configuration space). This aspect of DSR was certainly known but never exploited
before from the group field theory perspective. In fact having this in mind will allow us to derive
a DSR scalar field theory from a group field theory describing the BF quantum amplitudes in the
next section.

Before doing so, we recall the definition of the $\ka$-Minkowski
space and its associated momentum space, the $\AN$ group. The
construction can be done in any dimension. This means that we can
also obtain, in principle, an effective field theory on
$\ka$-Minkowski spacetime in any dimension from a group field
theory, using our procedure. However, we focus on the 4d case
which is directly relevant for quantum gravity. We then review the
construction of scalar field theory on $\ka$-Minkowski,
emphasizing the group field theory aspect.

\subsection{$\ka$-Minkowski and  the $\AN$ momentum space}

As a vector space, the $\ka$-Minkowski space-time is isomorphic to $\R^{n}$ and is defined as the
Lie algebra $\an_{n-2}$, which is  a subalgebra of the Lorentz algebra $\so(n-1,1)$. In the
following, we work with the signature $(-,+,...,+)$. The $n$-1 generators of $\an_{n-2}$ are given
by:
\be
X_0=\f{1}{\ka}J_{n0},\quad X_k=\f{1}{\ka}(J_{nk}+J_{0k}), \quad k=1,...,n-2,
\ee
where the $J_{\mn}$ are the generators of the Lorentz algebra $\so(n-1,1)$. It is easy to see that
$\an_{n-2}$ is therefore encoded by the following commutation relations:
\be \label{an3}
[X_0,X_k]\,=-\f{i}{\ka}X_k,\qquad [X_k,X_l]\,=0, \quad k,l=1,...,n.
\ee
Their explicit matrix elements in the fundamental  ($n$-dimensional) representation of $\so(n-1,1)$
are \cite{klymyk}:
\be
X_0=\f i\ka\mat{ccc}{0 & {\bf 0} & 1 \\ {\bf 0}&{\bf 0}&{\bf 0} \\ 1 & {\bf 0} &0},
\qquad
X_k=\f i\ka\mat{ccc}{0 & {}^t{\bf x} & 0 \\ {\bf x}&{\bf 0}&  {\bf  x}\\ 0 & -{}^t{\bf  x}
&0},
\ee
where ${}^t{\bf  x}$ are the $(n-2)$-dimensional basis vectors $(1,0,...,0)$, $(0,1,0,...)$, and so
on. For explicit calculations, it is convenient to notice that the matrices $X_k$ are nilpotent
with $(X_k)^3=0$. There are indeed $n$-2 abelian and nilpotent generators, hence the name
$\AN_{n-2}$. The corresponding exponentiated group elements are:
\be
e^{ik_0X_0}=\mat{ccc}{\cosh\f{k_0}{\ka} & {\bf 0} & -\sinh\f{k_0}{\ka} \\ {\bf
0}&{\mathbbm{1}}&{\bf 0},
\\ -\sinh\f{k_0}{\ka} & {\bf 0} &\cosh\f{k_0}{\ka}}
\qquad
e^{ik_iX_i}=\mat{ccc}{1+\f{{\bf k}^2}{2\ka^2} & -\f{{}^t{\bf k}}{\ka} & \f{{\bf k}^2}{2\ka^2} \\
-\f{{\bf k}}{\ka}&{\mathbbm{1}}&  -\f{{\bf k}}{\ka}\\
-\f{{\bf k}^2}{2\ka^2} & \f{{}^t{\bf k}}{\ka} &1-\f{{\bf k}^2}{2\ka^2}},
\ee
where $\mathbbm{1}$ is the $(n-2)\times(n-2)$ identity matrix. We parameterize generic $\AN_{n-2}$
group elements as
\be
\label{planewave} h(k_\mu)=h(k_0,k_i)\,\equiv\,e^{ik_0X_0}e^{ik_iX_i}.
\ee
As we will see in the next subsection, this group element can be interpreted as the non-commutative
plane-wave and the coordinates on the group $k_\mu$ as the wave-vector (and therefore related to
the momentum).
To multiply group elements in this parametrization, we check that:
$$
e^{ik_0X_0}e^{ik_iX_i}\,=\, e^{i(e^{{k_0}/\ka})k_iX_i}\,e^{ik_0X_0}.
$$
This is the exponentiated version of the commutation relation between $X_0$ and the $X_i$'s. This
allows to derive the multiplication law for $\AN_{n-2}$ group elements:
\be
h(k_0,{k_i})h(q_0,{ q_i})\,=\, h(k_0+q_0,e^{-q_0/\ka}{ k_i}+{ q_i}),
\ee
which defines a deformed non-commutative addition of the wave-vectors:
\be
(k \oplus q)_0 \,\equiv\, k_0+q_0,\qquad (k \oplus q)_i \,\equiv\, e^{-q_0/\ka}k_i+q_i.
\ee
This also gives the inverse group elements:
\be\label{addition-momentum}
h(k_0,k_i)^{-1}\,=\,h(-k_0,-e^{k_0/\ka}k_i),
\ee
which defines the opposite momentum $S(k_\mu)$ for the non-commutative addition:
\be
S(k_0)=-k_0, \qquad S(k_i)= -e^{k_0/\ka}k_i\label{inverse}.
\ee

The relation between the $\SO(n-1,1)$ group and $\AN_{n-2}$ is given by the Iwasawa decomposition
(see e.g. \cite{klymyk,iwasawa}):
\be
\SO(n-1,1)\,=\,
\AN_{n-2}\,\SO(n-2,1)\,\cup\,\AN_{n-2}\cM\,\SO(n-2,1),
\ee
where the two sets are disjoint and $\cM$ is the following diagonal matrix,
$$
\cM=\mat{ccc}{-1 & &  \\ &{\mathbbm{1}}& \\ &  &-1}.
$$

To understand the geometric meaning of this decomposition, we look at the map between $\AN_{n-2}$
and the de Sitter space-time $dS_{n-1}$ defined as the coset $\SO(n-1,1)/\SO(n-2,1)$. We introduce
a reference space-like vector $v^{(0)}\equiv(0,...,0,1)\in \R^n$. The little group of this vector
is  the Lorentz group $\SO(n-2,1)$ and the action of $\SO(n-1,1)$ on it sweeps the whole de Sitter
space.

Looking at the action of $\AN_{n-2}$ on $v^{(0)}$, using the plane-wave parametrization
\eqref{planewave}, we define the vector $v\,\equiv\, h(k_\mu).v^{(0)}$ with explicit coordinates:
\bes\label{5d momentum}
v_0&=& -\sinh\f{k_0}{\ka}+\f{{\bf k}^2}{2\ka^2}e^{k_0/\ka} \nn\\
v_i&=& -\f{k_i}{\ka} \\
v_n &=& \cosh\f{k_0}\ka -\f{{\bf k}^2}{2\ka^2}e^{k_0/\ka}. \nn
\label{5dparam}
\ees
We easily check that $v_Av^A=-v_0^2+\vec{v}^2+v_n^2=1$. However, since $v_0+v_n=\exp(-k_0/\ka)$,
this action of $\AN_{n-2}$ on $v^{(0)}$  sweeps only the half of the de Sitter space defined by the
condition $v_+=v_0+v_n>0$. Assuming this condition, we can reverse the previous relation and
express the $\AN_{n-2}$ group element in terms of the $n$-vector $v$:
\be
h(k_\mu)=\left(\begin{array}{ccc}v_n+\frac{{\bf  v}^2}{v_0+v_n}& \frac{{}^t\bf v}{v_0+v_n}&
v_0\\{\bf v}& \one & {\bf v}\\v_0-\frac{ {\bf  v}^2}{v_0+v_n}& \frac{-{}^t\bf v}{v_0+v_n}&
v_n\end{array}\right),
\qquad \textrm{with } h(k_\mu)^{-1}=\left(\begin{array}{ccc}v_n+\frac{{\bf  v}^2}{v_0+v_n}&
-{}^t\bf v& -v_0+\frac{ {\bf  v}^2}{v_0+v_n}\\\frac{-\bf v}{v_0+v_n}& \one & \frac{-\bf
v}{v_0+v_n}\\-v_0 & {}^t{\bf v}& v_n\end{array}\right).
\ee
To recover the full de Sitter space, we need to use the other part of the Iwasawa decomposition.
Considering the action of $\cM$, we obtain:
\be
h(k_\mu)\cM.v^{(0)}\,=\,-v\,\qquad h(k_\mu)=\left(\begin{array}{ccc}-v_n-\frac{{\bf v}^2}{v_0+v_n}&
\frac{{}^t\bf v}{v_0+v_n}& -v_0\\-{\bf v}& \one & -{\bf v}\\-v_0+\frac{ {\bf  v}^2}{v_0+v_n}&
\frac{-{}^t\bf v}{v_0+v_n}& -v_n\end{array}\right).
\ee
Thus the action of the $\cM$ operator simply maps the $n$-vector $v_A$ in its opposite $-v_A$.
Clearly that allows to complete the other side of de Sitter space with $v_+<0$. Let us point out
that the left action $\cM\,h(k_\mu)$ would still map $v_0\arr -v_0$ and $v_n\arr -v_N$ but would
leave the other components invariant ${\bf v}\arr {\bf v}$.

To summarize, an arbitrary point $v$ on the de Sitter space-time is uniquely obtained as:
\be
v\,=\,(-)^\eps h(k_\mu).v^{(0)}= h(k_\mu)\cM^\eps.v^{(0)},\qquad \eps=0\textrm{ or }1,\quad
h\in\AN_{n-2}.
\ee
The sign $(-)^\eps$ corresponds to the two components of the Iwasawa decomposition. The coset space
$\SO(n-1,1)/\SO(n-2,1)$ is isomorphic to the de Sitter space and is covered by two patches, each of
these patches being isomorphic to the group $\AN_{n-2}$.

We introduce the set $\AN^c_{n-2}\equiv\AN_{n-2}
\cup\AN_{n-2}\cM$, such that the Iwasawa decomposition reads
$\SO(n-1,1)=\AN^c\,\SO(n-2,1)$ and that $\AN^c$ is isomorphic to
the full de Sitter space (without any restriction on the sign of
$v_+$). Actually, $AN^c_{n-2}$ is itself a group. Indeed we first
easily check the commutation relation between the $\cM$ operator
and $\AN_{n-2}$ group elements:
$$
\cM h(k_\mu) = h(k_0,-k_i) \cM,
$$
where commuting $\cM$ with $h$ sends the 5-vector $v_A$ to
$(v_0,-{\bf v},v_n)$. This implies the group multiplication on
$\AN^c$: \be h(k_\mu)\cM^\alpha\,h(q_\mu)\cM^\beta\,=\, h(k\oplus
(-)^\alpha q) \cM^{\alpha+\beta}, \ee with $\alpha,\beta=0,1$.
Finally, we point out that $\AN^c_{n-2}$ is a group but not a Lie
group (because of the discrete $\Z_2$ component).

\medskip

In the following we will focus on $n=5$ case looking at $\SO(4,1)$
and its subgroup $\AN_3$ relevant for 4d Deformed Special
Relativity and quantum gravity. Consider the action of the Lorentz
transformations $\SO(3,1)$ on $AN_3$. This is not simple when seen
from the 4d perspective, i.e. from the point of view of $AN_3$
itself. However, it amounts to the obvious linear action of
$\SO(3,1)$ on the de Sitter space-time $dS_4$, $\Lambda\rhd
v\,=\,\Lambda.v$, leaving the fifth component $v_4$ invariant.
This leads to a non-linear action of $\Lambda\in\SO(3,1)$ on
$\AN_3$ (see e.g. \cite{iwasawa}): \be \Lambda\rhd
h(k_\mu)\cM^{\eps} \,\equiv\,\Lambda\, h(k_\mu)
\cM^{\eps}\,\tLambda^{-1} \,=\, h(k'_\mu)\cM^{\eps'}, \ee where
$\tLambda$, a priori different from $\Lambda$, is the unique
Lorentz transformation ensuring that the resulting group element
lives in $\AN^c_3\equiv\AN_3 \cup\AN_3\cM$. An important point is
that it is impossible to neglect the effect of $\cM$. Indeed the
Lorentz transformation mixes the two parts of the Iwasawa
decomposition: the subgroup $\AN_3$ is not invariant under the
$\SO(3,1)$ action but the group $\AN^c_3$ is.

It is possible to compute the ``counter-boost" $\tLambda$ for infinitesimal Lorentz transformations
\cite{iwasawa}. This leads to the $\ka$-Poincar\'e algebra presented as a non-linear realization of the Poicar\'e algebra in
terms of $k_\mu$:
\begin{eqnarray}\label{kappa poincare alegbra}
&& [M_i,k_j]=\epsilon_{ij}^lk_l, \quad [M_i,k_0]= 0, \quad [k_\mu,k_\nu]=0\nn\\
&& [N_i,k_j]= \delta_{ij}\left(\sinh\f{k_0}{\ka}-\f{{\bf k}^2}{2\ka^2}e^{k_0/\ka}\right), \quad [N_i,k_0]= k_i e^{k_0/\ka}
\end{eqnarray}

Finally, we will need an integration measure on $\AN_3$ in order to define a Fourier transform. The
group $\AN_3$ is provided with two invariant Haar measures:
\be
\int dh_L=\int d^4k_\mu,\quad  \int dh_R=\int e^{+3k_0/\ka}\,d^4k_\mu,
\ee
which are respectively invariant under the left and right action of the group $\AN_3$. Let us point
out that:
$$
\int d(h^{-1})_L\,=\,\int dh_R.
$$
We can easily derive this measure from the 5d perspective using the parametrization \Ref{5dparam}:
\be\label{measure5d}
\ka^4\int\delta(v_Av^A-1)\theta(v_0+v_4)\,d^5 v_A\,=\,\int d^4k_\mu\,=\,\int dh_L,
\ee
where the $\theta(v_+)$ function imposes the $v+>0$ restriction. Indeed the $\SO(4,1)$ action on
the reference vector $v^{(0)}$ generates the whole de Sitter space,
$$
v=g\rhd v^{(0)}=h\cM^\eps\Lambda\rhd v^{(0)}=h\cM^\eps\rhd v^{(0)}.
$$
Therefore the natural measure on $\AN_3$ inherited from the Haar measure on $\SO(4,1)$ is
left-invariant.

A crucial issue is the Lorentz invariance of the measure. Even
though the measure $dh_L=d^4k_\mu$ looks Lorentz invariant, it is
not, as the action of the Lorentz group on
the coordinates $k_\mu$ is non-trivial and non-linear. Actually,
one can show this action does not leave the measure invariant.
What causes the problem is the restriction $v_+>0$ (needed when
inducing the measure on $AN_3$ from the Lorentz invariant measure
on $dS_4$, which indeed breaks Lorentz invariance. In order to get
a Lorentz invariant measure, we need to glue back the two patches
$v_+<0$ and $v_+>0$ (and actually also the $v_+=0$ patch) and
define the measure on the whole de Sitter space. In other words,
we write the same measure as a measure on $\AN^c_3\equiv\AN_3
\cup\AN_3\cM\sim dS$: \be \int dh_L\equiv \int_{\AN_3} dh_L^+ +
\int_{\AN_3\cM} dh_L^-=\int\delta(v_Av^A-1)d^5v. \ee Another way
to circumvent this problem and obtain a Lorentz invariant measure
is to consider a space without boundary and work on the so-called
elliptic de Sitter space\footnotemark~$dS/\Z_2$ where we identify
$v_A \leftrightarrow -v_A$, which amounts to identifying the group
elements $h(k_\mu)\leftrightarrow h(k_\mu)\cM$. This space is
indeed isomorphic to $AN_3$ as a manifold. One way to achieve
nicely this restriction at the field theory level is to consider
only fields on de Sitter space (or on $AN_3^c$) which are
invariant under the parity transformation $v_A \leftrightarrow
-v_A$ \cite{FKG}. In this case, we recover the measure $d^4k_\mu$
on $\AN_3\sim \AN^c_3/\Z_2 \sim dS/\Z_2$.

\footnotetext{
Considering deformed special relativity in three dimensions with Euclidean signature, the group
field theory on $\SU(2)$ has a similar feature \cite{karim}. $\SU(2)$ being isomorphic to the
sphere $S^3$ is indeed also covered by two patches. Note however than in this case the standard
choice of coordinates is not breaking the Lorentz symmetries. To get rid of one patch, we identify
the two patches and consider instead $\SO(3)=\SU(2)/\Z_2$ as in
\cite{effqg,matrix}.}


\subsection{DSR Field Theory (in a nutshell)}
We now present a DSR  scalar field theory first as a group field
theory. Then we recall how we can recover the scalar field theory
on $\ka$-Minkowski using a generalized Fourier transform. For
simplicity, we shall restrict to the case $n=5$, so that we shall
consider the non-compact and non-semi-simple groups $G=AN^c_3,\,
AN_3$.

We consider the real scalar field $\phi: G\rightarrow \R$, and define the (free) action
\be
\label{action DSR 1}
\ss(\phi)= \int dh_L \, \phi(h)\, \kk(h)\, \phi(h), \quad \forall h\in G,
\ee
where $\kk(h)$ is the propagator and $dh_L$ is the left invariant measure. Contrary to the usual
group field theory philosophy, we interpret $G$ as the momentum space.

First let us discuss  the possible choices of propagators. We
demand $\kk(h)$ to be a function on $G$ invariant  under the
Lorentz transformations. We have showed in the previous subsection
how the Lorentz group is acting on $G$. It is then clear that any
function $\kk(h)=f(v_4(h))$ is  a good candidate, since $v_4$ is
by construction a Lorentz invariant quantity.  Two main choices
have been studied in the literature. \be \kk_1(h)= (\kappa
^2-\pi_4(h)) -m^2, \quad \kk_2(h)= \kappa ^2-(\pi_4(h))^2-m^2,
\quad \pi_4=\kappa v_4. \ee The freedom in choosing the propagator
is related to the ambiguity in choosing what we call momentum. To
have a precise candidate for the notion of momentum, one needs to
define first position and define momentum either as the eigenvalue
of the translation operator applied to the plane-wave and/or the
conserved charged for the action $\ss(\phi)$ expressed in terms of
coordinates associated to the translations \cite{FKG,michele}.
Therefore from the group field theory perspective, it is necessary
to perform a Fourier transform to obtain more information.

Before introducing the Fourier transform, let us note that the
action \eqref{action DSR 1} is clearly Lorentz invariant if the
measure is  Lorentz invariant, since the propagator $\kk(h)$ has
been chosen to be a Lorentz invariant function and the
transformation of the fields induced by a Lorentz transformation
on the arguments $h$ is also known, from the previous subsection.
We have also seen that this measure is indeed a Lorentz invariant
measure both in the case of group manifold $AN_3^c$ and generic
scalar fields, and in the case of elliptic de Sitter space or
$AN_3$ when a restriction to symmetric fields is imposed.



The generalized Fourier transform relates functions on the group $\cc(G)$ and elements of the
enveloping algebra $\uu(\an_3)$. It is defined respectively for $G=\AN^c_3, \AN_3$ as
\begin{eqnarray}
\label{fourier}
&&\hat \phi(X)= \int_{\AN_3} dh_L^+ \, h(k_\mu)\, \phi^+(k)+\int_{\AN_3\cM} dh_L^- \, h(k_\mu)\, \phi^-(k), \quad X\in \an_3,
\quad \hat \phi(X) \in \uu(\an_3)\\
&&\hat \phi(X)= \int_{\AN_3} dh_L \, h(k_\mu)\, \phi(k), \quad X\in \an_3, \quad \hat \phi(X) \in \uu(\an_3)
\end{eqnarray}
where we used the non-abelian plane-wave $h(k_\mu)$. The inverse
Fourier transform can also be introduced, if one introduces a
measure $d^4X$. For the details we refer to \cite{majid,FKG}. The
group field theory action on $G$ can now be rewritten as a
non-commutative field theory on $\ka$-Minkowski (to simplify the
notation we restrict our attention to  $G=\AN_3$ and thus we
implicitly consider symmetric fields). \be \label{action DSR 2}
\ss(\phi)= \int dh_L \, \phi(h)\, \kk(h)\, \phi(h) = \int \,
d^4X\, \left(
\partial_\mu\hat\phi(X)
\partial^\mu\hat\phi(X) + m^2 \hat\phi^2(X)\right). \ee The
Poincar\'e symmetries are naturally deformed in order to be
consistent with the non-trivial commutation relations of the
$\ka$-Minkowski coordinates. More exactly, if the Poincar\'e
transformations act in the standard on the
coordinates\footnote{$T_\mu, N_i, R_i$ are respectively
translations,  boosts and rotations.}
\begin{eqnarray}
&& T_\mu\act X_\nu =\delta_\mn,  \quad  N_i\act X_j= \delta_{ij}X_0, \quad N_i\act X_0= X_i \nn \\
&& R_i \act X_j= \epsilon_{ij}^k X_k, \quad  R_i \act X_0= 0,
\end{eqnarray}
its action on the product of coordinates has to be modified in order to be consistent with the
non-trivial commutation relation \eqref{an3}, that is we demand that
$$ \ttt \act [X_\mu,X_\nu]=  C_{\mn}^\alpha \,\ttt\act X_\alpha, \quad \forall \ttt= T_\mu, R_i, N_i,$$
and $C_{\mn}^\alpha$ is the structure constant of $\an_3$. To implement this one needs to deform
the coalgebra structure of the Poincar\'e algebra, that is one deforms the
coproduct\footnote{Indeed, we have for example for a translation
\be
\label{consistency}
T_\mu\act (X_\alpha X_\beta)=  T_\mu\act m (X_\alpha \otimes X_\beta)= m[(\cop T_\mu)\act (X_\alpha \otimes X_\beta)],
\ee
where $m$ is the multiplication.} $\cop$
\begin{eqnarray}
\label{coprod}
&& \cop T_\mu = T_\mu\otimes\one + \one\otimes T_\mu - \ka\mone T_0 \otimes T_\mu  \nn\\
&& \cop N_i=  N_i\otimes\one + \one\otimes N_i  - \ka\mone T_0\otimes N_i + \ka\mone{\epsilon_{i}}^{jk} T_k \otimes R_j\nn\\
&& \cop R_i = R_i\otimes\one + \one\otimes R_i.
\end{eqnarray}
Thanks to this new coproduct,  the Poincar\'e transformations and the commutation relations
\eqref{an3} are consistent, ie \eqref{consistency} is true. Moreover, using the coproduct, we can
act on the plane-wave and deduce the realization of the Poincar\'e transformations in terms of the
coordinates $k_\mu$. We recover precisely the $\ka$-algebra \eqref{kappa poincare alegbra} as one
could have guessed. Finally, as we mentioned earlier, the "physical" notion of momentum $\pi_\mu$
can be identified from the action of the translations on the plane-wave
$$T_\mu\act h(k_\nu)\equiv \pi_\mu\, h(k_\nu).$$
Direct calculation \cite{FKG}, using again the coproduct shows
that $$\pi_\mu=\kappa v_\mu.$$ We have therefore a non-linear
relation between the wave-vector $k_\mu$ and the momentum
$\pi_\mu$. Moreover, using the 5d bicovariant differential
calculus, it was also shown  that the conserved charges,   for the
free action \eqref{action DSR 2},   associated to the translations
are precisely $\pi_\mu$  \cite{FKG}. With this choice of momentum
the propagator $\kk_2(h)$ becomes simply $\kk_2(h)=\pi_\mu\pi^\mu
-m^2$, thanks to the de Sitter constraint $\pi_A\pi^A=\kappa^2$.


\section{Deriving Deformed Special Relativity from Group Field Theory}
We now come to the main issue we address in this paper: to obtain
a field theory on $\ka$-Minkowski (or equivalently on $AN_3$
momentum space) from a 4d group field theory, in particular from
one that could be related to 4d Quantum Gravity. We have already
shown the general construction leading from a generic 4d GFT to an
effective QFT based on the same group manifold. Now the task is to
specialize that construction to the case of physical interest.

We start from the group field theory describing topological
BF-theory for the non-compact gauge group $\SO(4,1)$.

There are several reasons of interest in this model. First of all,
the McDowell-Mansouri formulation (as well as related ones
\cite{artem}) of General Relativity with cosmological constant
defines 4d gravity as a BF-theory for $\SO(4,1)$ plus a potential
term which breaks the gauge symmetry from $\SO(4,1)$ down to the
Lorentz group $\SO(3,1)$. On the one hand, this leads to the idea
of understanding gravity as a phase of a fundamental topological
field theory, an idea that has been put forward several times in
the past. On the other hand, it suggests to try to define Quantum
Gravity in the spin foam context as a perturbation of a
topological spin foam model for $SO(4,1)$ BF theory. These ideas
could also be implemented directly at the GFT level. If one does
so, the starting point would necessarily be a GFT for $SO(4,1)$ of
the type we use below. Second, as this model describes $SO(4,1)$
BF theory in a \lq\lq 3rd quantized\rq\rq setting, we expect any
classical solution of the GFT equations to represent quantum de Sitter space on some given topology, analogously to what happens
with Minkowski space in the $SO(3,1)$ case. Such configurations
would most likely be present (and physically relevant) also in a
complete non-topological gravity model obtained starting from the
topological one. Third, and partly as a consequence of the above,
to start from the spinfoam/GFT model for $\SO(4,1)$ BF-theory
seems to be the correct arena to build a spin foam model for 4d
quantum gravity plus particles on de Sitter space \cite{mm},
treating particles as arising from topological curvature defects
for an $SO(4,1)$ connection, along the lines of what has been
already achieved in 3d gravity \cite{effqg}.

We do not describe the structure of the corresponding spin foam
path integral, as the spin foam (perturbative) formulation plays
no role in our construction. We start instead directly with the
relevant group field theory, and work only at the level of the GFT
action. As in the compact group case, we consider a gauge
invariant field on $\SO(4,1)^{\times 4}$:
$$
\phi(g_1,g_2,g_3,g_4)\,=\,\phi(g_1g,g_2g,g_3g,g_4g), \quad \forall g\in\SO(4,1),
$$
and the group field action is given by: \bes S_{4d}&=&\f12\int
[dg]^3\,\phi(g_1,g_2,g_3,g_4)\phi(g_4,g_3,g_2,g_1) \\&&
-\f{\lambda}{5!} \int [dg]^{9}
\phi(g_1,g_2,g_3,g_4)\phi(g_4,g_5,g_6,g_7)\phi(g_7,g_3,g_8,g_9)\phi(g_9,g_6,g_2,g_{10})\phi(g_{10},g_8,g_5,g_1).
\label{so41} \ees Because of the symmetry requirement, one of the
field arguments in redundant, and one can effectively work with a
field depending on only three group elements. This is indicated
schematically above, where, we integrate only over three group
elements in the kinetic term and nine in the interaction term in
order to avoid redundant integrations, which would lead to
divergences due to the non-compactness of the group $SO(4,1)$.
More precisely, considering the kinetic term, we can fix one of
the four group elements, say $g_4$, to an arbitrary value (usually
the identity $\id$) and integrate over the remaining three group
elements without changing anything to the final result. Similarly,
the restriction to only nine integrations in the interaction term
can be understood as a partial gauge fixing, avoiding redundancies
and associated divergences.

Starting with this group field theory, we want to derive the DSR
field theory as a sector of the full theory. We follow the same
strategy as in the three-dimensional case and as outlined earlier
for the 4-dimensional case: we search for classical solutions of
the $\SO(4,1)$ group field theory and study specific
two-dimensional field variations around it. We will naturally
obtain an effective field theory living on $\SO(4,1)$. On top of
this, we want then to obtain, from such effective field theory,
one that is restricted to the $AN_3^3$ (or $AN_3$) homogeneous
space (subgroup). There are three main strategies following which
this could be achieved, a priori:
\begin{itemize}
\item We could derive first an effective field theory on
$\SO(4,1)$ and then study the possibility and mechanism for a
decoupling of the $AN_3^c$ degrees of freedom from the ones living
on the Lorentz $SO(3,1)$ sector of the initial $SO(4,1)$ group.

\item We could try to identify some special classical solutions of
the fundamental $SO(4,1)$ group field theory, which are such that
the effective matter field would naturally result in being
localized on $\AN^c_3$.

\item We could modify the initial $SO(4,1)$ group field theory
action in such a way that, after the same procedure, the resulting
effective matter field is automatically localized on $\AN^c_3$ (or
$\AN_3$).

\end{itemize}
Anticipating the results of this section, we will see that the
first strategy leads naturally to a DSR kinetic term, depending
only on $AN_3$ degrees of freedom, and thus with an exact
decoupling of the $SO(3,1)$ modes. However, the interaction term
still couples the two sets of modes. This leads to the suggestion
that the reduction to a pure $AN_3$ theory can be a dynamical
effect, and we will show the effective pure $AN_3$ theory. As for
the second strategy, we will see that it does not work as simply
as stated, and it requires necessarily a modification of the
initial group field theory action, i.e. to some version of the
third strategy. We will discuss some ways in which this can be
implemented, but we will see that the simplest way to achieve this
is to start directly with a group field theory for BF-theory with
gauge group $\AN_3^c$.

\subsubsection{Deformed Special Relativity as a Phase of SO(4,1) GFT}

Let us start from the action above defining the group field theory for the $\SO(4,1)$ BF-theory.
The first task is to write the field equations and identify classical solutions. This works as in
the compact group case presented in section \ref{compact}. We use the same ansatz:
$$
\phi^{(0)}(g_i)\,=\,{}^3\sqrt{\f{4!}{\lambda}}\,\int_{\SO(4,1)}
dg\,\delta(g_1g)F(g_2g)\tF(g_3g)\delta(g_4g),
$$
where the functions $F$ and $\tF$ must satisfy the normalization condition $\int F\tF=1$. Moreover,
we also require that $\int F$ and $\int \tF$ be finite in order to get a meaningful effective
action for the 2d field variations around the classical solutions.

The ansatz that we choose is tailored to lead us to the DSR field
theory~\footnotemark~: \be F(g)\,=\, \alpha(v_4(g)+a)\vartheta(g),
\qquad \tF(g)\,=\,\delta(g). \ee The function $v_4$ is defined as
matrix element of $g$ in the fundamental (non-unitary)
five-dimensional representation of $SO(4,1)$, $v_4(g)\,=\,\la
v^{(0)}|g|v^{(0)}\ra$, where $v^{(0)}=(0,0,0,0,1)$ is, as
previously, the vector invariant under the $SO(3,1)$ Lorentz
subgroup. $\vartheta(g)$ is a cut-off function providing a
regularization of $F$, so that it becomes an $L^1$ function. We
first check the normalization condition  $\int
F\tF=\alpha(a+1)\vartheta(\id)=1$, and, assuming that
$\vartheta(\id)=1$, we require $\alpha=(a+1)^{-1}$ in order for it
to be satisfied.

\footnotetext{ We can also choose a more symmetric ansatz with
$F(g)=\tF(g)$ which would correspond to a group field satisfying
the reality condition. The resulting calculations would be more
involved, and this is why we do not discuss in detail this choice.
However, it can be easily checked that, with a similar
regularization, the final result would be the same.}

Then we can derive the effective action around such classical
solutions for 2d field variations just as in the compact group
case given in \eqref{compacteffaction}: \bes S_{eff}[\psi]&=&
\f12\int
\psi(g)\psi(g^{-1})\left[1-2c^2-\vartheta^2(g)\f{2c(a+v_4(g))^2}{(a+1)^2}\right]
-c\left({}^3\sqrt{\f{\lambda}{4!}}\right)\,\int\psi(g_1)..\psi(g_3)\,\delta(g_1..g_3)
\left[c+F(g_3)\right] \nn\\ &&
-c\left({}^3\sqrt{\f{\lambda}{4!}}\right)^2\,\int
\psi(g_1)..\psi(g_4)\,\delta(g_1..g_4) -\f\lambda{5!}\int
\psi(g_1)..\psi(g_5)\,\delta(g_1..g_5),
\label{noncompacteffaction} \ees where $c=\int F$. Thus the last
issue to address in order to properly define this action is to
compute the integral of $F$. The function $v_4(g)$ is invariant
under the Lorentz group $\SO(3,1)$. Using the Iwasawa
decomposition $g=h\Lambda$ with $h\in AN^c_3$ and
$\Lambda\in\SO(3,1)$, it is easy to see that the matrix element
$v_4(g)$ actually only depends on $h$. Therefore it is natural to
split the cut-off function $\vartheta(g)$ in factors independently
regularizing the integrals over $\AN^c_3$ and over $\SO(3,1)$: \be
\vartheta(g)\,=\, \chi(h)\theta(\Lambda). \ee To keep calculations
simple, we assume that we choose the function $\theta(\Lambda)$ to
be a Gaussian function, or any other function peaked on
$\Lambda=\id$, such that $\theta(\id)=1$ and $\int \theta =1$.
Then using the isomorphism between $AN^c_3$ and the de Sitter
space $v_Av^A=1$, we choose the cut-off function on $\AN^c_3$ to
be $L^1$ and symmetric under $v_4\leftrightarrow-v_4$: the
simplest choice is to bound $|v_0|\le V$, which automatically also
bounds $v_4$ and ${\bf v}$. We get: \be c=\int F = \int dh
\chi(h)\,\f{a+v_4(h)}{a+1} = \int [d^5v_A]\,\delta(v_4^2+{\bf
v}^2-v_0^2-1)\, \chi(v_A)\f{a+v_4}{a+1} = \f{a\int_{dS}
\chi}{a+1}, \ee since $v_4$ is a odd function on the de Sitter
space. For our simplest choice of $\chi$-function imposing a
straightforward bound on $v_0$, we easily evaluate: \be \int_{dS}
\chi(v)\,=\,4\pi\int_{-V}^{V} dv_0
\int_{-\sqrt{1+v_0^2}}^{\sqrt{1+v_0^2}} dv_4 \sqrt{1+v_0^2-v_4^2}
\,=\, \f{4\pi^2}{3}\,V(V^2+3). \ee For more generic choices of
cut-off functions $\chi$, the last factor $\int_{dS}\chi$ is at
most quartic\footnotemark{} in the cut-off value $V$.

If we want to remove the cut-off and re-absorb all the infinities
due to the non-compactness of the group, we could now send the
cut-off $V$ to $\infty$, and then we also send the factor $a$ to
0, scaling it as $a\propto 1/V^3$. In this way, we keep $c$
finite. This is the simplest method to achieve the result, but of
course others can be considered. We point out that this
renormalisation is done at the classical level in the definition
of our classical solution and not at the quantum level like in
quantum field theory. In other words, this regularization is
necessary in order to obtain a true and well-defined classical
solution of the equations of motion, and meaningful variations
around it.

\footnotetext{
As an example, for a cut-off function $\chi$ implementing directly a bound on $v_4$ and the
3-vector ${\bf v}$, we have:
$$
\int_{dS} \chi(v)\,=\,4\pi\int_{-V}^{+V}dv_4\int^{+V}dv\, \f{v^2}{2\sqrt{v^2+v_4^2-1}} \propto V^3\ln V.
$$ }

After all these regularization details, in the double scaling
limit\footnotemark{} $a\arr0$ and $L\arr\infty$ while keeping $c$
finite, we have derived an effective theory for a field $\psi(g)$
living on $\SO(4,1)$: \bes S_{eff}[\psi]&=& \f12\int
\psi(g)\psi(g^{-1})\left[1-2c^2-2cv_4(h)^2\chi(h)^2\theta(\Lambda)^2\right]
-{}^3\sqrt{\f{\lambda}{4!}}\,\int\psi(g_1)..\psi(g_3)\,\delta(g_1..g_3)
\left[c^2+cF(g_3)\right] \nn\\ &&
-c\left({}^3\sqrt{\f{\lambda}{4!}}\right)^2\,\int
\psi(g_1)..\psi(g_4)\,\delta(g_1..g_4) -\f\lambda{5!}\int
\psi(g_1)..\psi(g_5)\,\delta(g_1..g_5). \ees We recognize the
correct kinetic term for a DSR field theory. However, the
effective matter field is a priori still defined on the full
$SO(4,1)$ momentum manifold. The only remaining issue is therefore
to understand the ``localization" process of the field $\psi$ to
$\AN^c_3$. Having done this, we would truly have derived a scalar
field theory in deformed special relativity from the group field
theory defining topological $\SO(4,1)$ BF-theory, and thus a
sector of 4d quantum gravity.

\footnotetext{ Obviously, we do not {\it need} to take the limit.
We could keep $a,L,c$ all finite and define a solution
parameterized by these constants. As a result, we would simply get
extra constant terms in the action, e.g. terms in $a^2$ and
$av_4(h)$ in the propagator. The limiting procedure is implemented
only in order to get a simpler form of the action.}

Let us consider the second strategy envisaged above. A possible
solution to the localization issue is, the strategy goes, to use
the classical solution $F$ itself to localize the field on the
$AN^c_3$ manifold. For example, one may require that the
regularizing function $\theta(\Lambda)$ forces the $\SO(3,1)$
group element to be, say, the identity element, $\Lambda=\id$. The
simplest choice is to use a delta function on $SO(3,1)$. This
however causes two problems. First, both $\theta(\Lambda)$ and
$\theta(\Lambda)^2$ appear in the action above, and of course the
square of the $\delta$-distribution is not well-defined. One can
devise methods to overcome this purely mathematical problem, by
using suitable \lq\lq smoothed\rq\rq delta distributions, which
achieve the same localization, but are $L^2$ functions. The second
problem is however more fundamental. By construction, this method
forces the group element $g$ to lay in $\AN^c_3$ only in the terms
containing some factors $F(g)$, i.e. depending in a non-trivial
way (not as an overall constant) on the classical solution chosen.
Thus the mass term and most of the interaction terms are
completely transparent to this way of projecting on $\AN^c_3$. We
conclude that it is not enough to use the classical solution to
achieve this reduction from the full group $\SO(4,1)$ to the
sub-manifold $\AN^c_3$.

\medskip

We then look more carefully at the first strategy outlined above.
We see immediately that the kinetic term (containing the
differential operator defining the propagation of the field
degrees of freedom, as well as the symplectic structure in a
canonical setting, does not show any dependence on the Lorentz
sector. Indeed, through our choice of classical solution, we
obtained a kinetic term in $v_4(g)$ which depends only on the
$\AN^c_3$ part $h$ of the group element $g=h\Lambda$. This
suggests that the $\SO(3,1)$ degrees of freedom are non-dynamical
and that the restriction of the domain of defintion of the field
$\psi$ to $\AN^c_3$ group elements defines a dynamically stable
phase of the theory. This would be trivially true if not for the
fact that the interaction term does, a priori, depend also on the
Lorentz degrees of freedom, and couples them among the different
interacting fields. One way to make this manifest is, for example,
to assume that the perturbation field $\psi$ has a product
structure $\psi(g)=\tilde{\psi}(h)\Psi(\Lambda)$. We see that, as
far as the kinetic term is concerned, the only contribution from
the Lorentz sector is a constant multiplicative term
$\int_{SO(3,1)}d\Lambda\Psi(\Lambda)\Psi(\Lambda)$. Therefore we
get an exactly DSR-like and $\ka$-Poinca\'e invariant free field
theory. On the other hand, the vertex term couples Lorentz and
$AN_3$ degrees of freedom, and thus the $\tilde{\psi}$ and $\Psi$
parts of the field $\phi$; thus the $\ka$-Poincar\'e symmetry is
broken and the pure DSR-like form lost.

The above also shows that, if we were to choose the dependence of
the perturbation field on the Lorentz sector to be trivial, i.e.
$\Psi(\Lambda)\equiv 1$, and thus to {\it start} from a
perturbation field defined only on the $AN_3$ subgroup, we would
indeed obtain a nice DSR field theory, but with an interaction
term that would be more complicated that a simple polynomial
interaction. This would be due to to the integrations over the
Lorentz group manifold that, through the delta function, would
complicated the couplings between the $AN_3$ variables on which
the fields depend. Of course, it would be a possible DSR field
theory nevertheless.

\medskip

Still, because of the form of the kinetic term, we conjecture a
reduction to the $\AN_3$ sector to happen dynamically. This
dynamical reduction could be obtained in two main ways.\\ 
First, one could
expect that transition (scattering) amplitudes involving only real
particles defined on $\AN_3^c$, i.e. with momenta in this
submanifold, will not lead to creation of particles with Lorentz
degrees of freedom as well, due to the form of the propagator,
even if in principle they would be allowed by the enlarged
momentum conservation law coming from the interaction term, which
is defined on the full $\SO(4,1)$ group.\\
A second possibility is
that a proper canonical analysis of the effective field theory we
have obtained would show that the $\SO(3,1)$ modes are pure gauge
and can simply fixed from the start and thus drop from the action
altogether. We leave a more detailed analysis of this issue for
future work. Whether or not the restriction to $\AN_3$ is obtained
automatically, in one of the above ways, or by some other
procedure that will be revealed by a more detailed analysis, what
is certain is that a restricted theory obtained from the above and
living on $AN_3^c$ only is dynamically stable. In fact, if we
consider only excitations of the field in $\AN^c_3$, we will never
obtain excitations in $\SO(3,1)$ due to momentum conservation
$\delta(g_1..g_n)$ since $\AN^c_3$ is a subgroup. Therefore
$\AN^c_3$ is stable under the dynamics of the field theory, and
thus a restriction to fields on $\AN^c_3$ is consistent.

One can compare this situation to the case of a 2d field theory
written in momentum space where the propagator depends on $p_x$
and not on $p_y$:
$$
S_{eg}[\psi]=\int d^2\vec{p}\, (-p_x^2)\psi(\vec{p})\psi(-\vec{p}) +\int
[d^2p]^n\,\delta\left(\sum_i^n p^{(i)}\right)\prod_i^n\psi(p^{(i)}).
$$
The momentum $p_y$ does not enter the propagator and it defines a
pure gauge degree of freedom, as it can be checked by
straightforward canonical analysis. Therefore, we can restrict
ourselves to the sector $p_y=0$ without affecting the dynamics of
the field, nor any physical content of the theory.

Then, restricting ourselves to group elements $g_i=h_i\in\AN^c_3$,
we have the (non-commutative) field theory: \bes
S_{final}[\psi]&=& \f12\int dh_L
\psi(h)\psi(h^{-1})\left[1-2c^2-2cv_4(h)^2\chi(h)^2\right]
-{}^3\sqrt{\f{\lambda}{4!}}\,\int\psi(h_1)..\psi(h_3)\,\delta(h_1..h_3)
\left[c^2+cv_4(h_3)\chi(h_3)\right] \nn\\ &&
-c\left({}^3\sqrt{\f{\lambda}{4!}}\right)^2\,\int
\psi(h_1)..\psi(h_4)\,\delta(h_1..h_4) -\f\lambda{5!}\int
\psi(h_1)..\psi(h_5)\,\delta(h_1..h_5), \label{finaldsraction}
\ees where the left-invariant measure on $\AN^c_3$ is inherited
from the Haar measure on $\SO(4,1)$. We argue that this is the
theory that encodes the full dynamics of 2d perturbations, as
emergent matter fields, of the $SO(4,1)$ GFT, around the special
classical solution we have chosen. We have thus finally derived
the scalar field theory for deformed special relativity with a
$\ka$-deformed Poincar\'e symmetry from the $\SO(4,1)$ group field
theory defining the transition amplitudes for the topological
BF-theory. To summarize, this was achieved in three steps:
\begin{enumerate}
\item Identify the correct regularized classical solution(s) to
the initial group field theory.

\item Look at the two-dimensional field variations around such a
classical solution and write the effective action describing their
dynamics.

\item Localize the field variations on the $\AN^c_3$ group
manifold relevant to deformed special relativity.
\end{enumerate}

An important remark is that we have a field theory already with an
in-built cut-off in momentum space due to the regularizing
function $\chi(h)$, necessary to define the classical solution to
the group field theory. Of course we can always send this cut-off
to infinity by the double scaling limit $V\arr\infty, a\arr 0$. At
the quantum level, we would anyway have to introduce such a
momentum cut-off to define the perturbative expansion of the
quantum field theory in term of Feynman diagrams. Here, on the
other hand, the momentum cut-off is not included to regularize the
Feynman diagrams, i.e. the discrete quantum histories of the
theory, but it appears naturally in our derivation of the
effective field theory on $\AN^c_3$ from the initial group field
theory on $\SO(4,1)$. Indeed, we insist on the fact that the
classical solution around which we study the group field variation
can not be defined without this momentum cut-off.

We conclude this case mentioning the open issue of the role of the
$\SO(3,1)$ degrees of freedom. As we noticed above, they would
actually couple non-trivially to the $\AN^c_3$ momentum, within
the full $\SO(4,1)$ theory, through the non-abelian momentum
conservation
$\delta(g_1g_2..)=\delta(h_1\Lambda_1h_2\Lambda_2..)$. We argued
that they may correspond to pure gauge degrees of freedom, and
consequently restricted the model to the pure $\AN_3^c$ sector.
Another possibility may be to use them to take into account spin
degrees of freedom, and thus define an emergent matter field
theory for spinning particles. In general, if it turns out that
our conjecture is incorrect, it would be interesting to show what
is their physical effect.

\subsubsection{Starting from a restricted GFT: the $\AN_3$ case}
Having followed in detail the first strategy outlined above, and
having shown the non-viability of the second, we now describe the
third, and obtain a DSR-like field theory in a different way.
Accordingly, instead of localizing the field variation on
$\AN^c_3$ in the final step, having first derived an effective
field theory on $\SO(4,1)$, we could modify our starting group
field theory action in such a way that the effective field theory
for perturbations is automatically localized on $\AN_3^c$.

The first case we deal with is the simplest one in which we choose
our initial fundamental theory to be itself a group field theory
for 4d BF theory with $\AN_3^c$ gauge group. We can then perform
the same analysis as in section \ref{4dcase}, and then choose the
same classical solution we have used in the previous section (now
seen as a function on the $\AN_3^c$ subgroup of $\SO(4,1)$ only.
This naturally to a field theory on $\AN^c_3$ describing a scalar
field with a deformed special relativity kinematics. The drawback
is that the link with 4d quantum gravity is now more obscure. It
is still possible that such group field theory is related to the
quantization of the McDowell-Mansouri formulation of 4d gravity,
but the exact relation is unclear. It still defines a topological
spin foam model, thus lacking any local gravity degree of freedom;
moreover, it lacks the information contained in $\SO(4,1)$ BF,
e.g. the cosmological constant, and its classical solutions have
no immediate spacetime interpretation, contrary to that case.

Still, it represents the easiest route to a DSR field theory from
GFT. The only issue that one has to be careful with in this case
is the question of the measure since we have to decide whether to
use the left or right invariant measure. Since the left-invariant
measure is the one inherited from the Haar measure on $\SO(4,1)$,
it seems to be the natural one to use. As before, we introduce the
gauge invariant group field on $(\AN^c_3)^{\times 4}$~:
$$
\phi(h_1,..,h_4)\,=\,\phi(h_1h^{-1},..,h_4h^{-1}),\qquad\forall h\in\AN^c_3,
$$
and the corresponding action:
\bes
S_{an}[\phi]&=&\f12\int [dh]^3\,\phi(h_1,h_2,h_3,h_4)\phi(h_4,h_3,h_2,h_1) \nn\\
&& -\f{\lambda}{5!} \int [dh]^{9}
\phi(h_1,h_2,h_3,h_4)\phi(h_4,h_5,h_6,h_7)\phi(h_7,h_3,h_8,h_9)\phi(h_9,h_6,h_2,h_{10})\phi(h_{10},h_8,h_5,h_1),
\nn \ees where we have used everywhere the left-invariant measure
$dh_L$ on $\AN_3^c$. As before we check that the ``flat solution"
ansatz, \be \phi^{(0)}\,\equiv\, {}^3\sqrt{\f{4!}{\lambda}}\,\int
dh_{(L)}
\delta(h_1h^{-1})F(h_2h^{-1})\tF(h_3h^{-1})\delta(h_4h^{-1}), \ee
provides a classical solution to the group field theory as soon as
$\int F\tF=1$. Thus we should choose the same ansatz for the arbitrary
functions: \be F(h)\,=\, \f{v_4(h)+a}{a+1}\chi(h), \qquad
\tF(h)\,=\,\delta(h), \ee where we choose exactly the same
regularizing function $\chi(h)$ as in the previous section, e.g.
the one imposing the bound $v_0(h)^2\le V^2$. We then look at the
effective action for two-dimensional field variation
$\phi^{(0)}+\psi(h_1h_4^{-1})$ around the classical solution.
Using the left-invariance of the measure, we end up of course with
the same effective scalar field theory \eqref{finaldsraction} living
on $\AN^c_3$.

As said, this gives the shortest path from a four-dimensional
group field theory and deformed special relativity. The natural
question in this context is nevertheless the physical
meaning/relevance of 4d BF-theory with gauge group $\AN^c_3$, from a 4d quantum gravity standpoint, as
we discussed.

\medskip

We may attempt to strengthen the link with the $\SO(4,1)$ theory,
while remaining within the third strategy outlined above, and go
beyond the purely $\AN^c_3$ case. In fact, there seems to be the
possibility of intermediate schemes: we could start from a {\it
restricted} GFT on $\SO(4,1)$, i.e. one based on the action
\ref{so41} but with some appropriate modification of the kinetic
and vertex terms tailored to give an $\AN_3$ effective theory for
the perturbations. One could hope also that this restricted GFT
would still be close to a group field theory quantization of the
McDowell-Mansouri action for gravity. We mention here only one
such restriction, leaving a more detailed study for future work:
instead of projecting all the group elements of the group field
$\phi(g_1,g_2,g_3,g_4)$ on the $\AN^c_3$ subgroup, we could simply
constrain the group elements $g_1$ and $g_4$ to lay in this
submanifold. It is easy to see that this automatically localizes
the field variations $\psi(g_1g_4^{-1})$ on $\AN^c_3$, and thus,
in the end, a DSR field theory, by the same procedure explained
above. The physical content of such modified GFT remains unclear,
however. It seems to us that issue of the physical meaning of a
group field theory partially or totally projected on $\AN^c_3$ can
only come from a more thorough study of the spinfoam/GFT
quantization of the McDowell-Mansouri reformulation of General
Relativity \cite{mm,artem}. Therefore, we do not discuss any such
mixed scheme any further.


\section{GFT vs emergent analog gravity models}
We conclude by pointing out the similarity of the procedure we have used in our GFT context, for deriving effective matter field theories, with the one that is customary in condensed matter analog gravity models, for deriving effective field theories for quasi-particles on emergent geometries. Indeed, we can see that they are fully analogous and that the differences are mostly of a purely technical nature.
In fact, GFTs have been argued to be a natural framework to discuss about
"emergent gravity" ideas \cite{daniele}, and in particular to use ideas from statistical physics and condensed matter theory to bridge the gap between the microscopic description of quantum spacetime provided by GFTs and the continuum physics we are accustomed to. Our results can be therefore motivated also from this perspective, and acquire thus an additional reason of interest on top of those mentioned in the introduction (this aspect is also discussed in \cite{danieleemergent}). Here we limit ourselves to presenting the main points of this perspective, and to showing in which sense our procedure resembles the one used in the context of condensed matter analog gravity models, and where it differs from it.

Let us first of all clarify what we mean by emergence. We call \lq\lq emergent\rq\rq some degrees of freedom which are only defined in a
given regime, and there in terms of more fundamental degrees of freedom. For
example, emergent degrees of freedom can be perturbations around some given
vacuum state, like in our GFT results, or collective degrees of freedom. In general, the classical theories for these emergent degrees of freedom only give effective theories upon quantization, in other words their quantum counterpart would be meaningful only in a limited regime. A complete quantization procedure can therefore take place only on the fundamental degrees of freedom.
A symmetry is called \lq\lq emergent\rq\rq if it applies to emergent degrees of
freedom only and thus is valid only in the same limited regime in which they can be consistently defined. In general,
the emergent symmetry is not related to nor part of the symmetries of the fundamental system. Moreover, if the emergent symmetry is not already among the
fundamental symmetries, then it is never exact but it is realized
only approximately in the effective theory.

Condensed matter theory offers several specific examples of systems in which the
collective behavior of the microscopic constituents in some
hydrodynamic approximation gives rise to effective emergent
geometries as well as matter fields (superfluid Helium-3, Bose-Einstein condensates, etc) \cite{analog-review}. Thus it supports the idea that
gravity and matter may emerge from
fundamental systems which do not have a geometric or gravitational
nature per se, at least in their fundamental formulation (like GFTs).

The emergence of gravity and (generically) curved geometries in
analog condensed matter models takes place usually in the hydrodynamic regime, which in turn is often
obtained in some mean field approximation around a background configuration
of the fluid. What happens is that
the collective parameters describing the fluid and its dynamics in
these background configurations (e.g. the density and velocity of
the fluid in the laboratory frame) can be recast as the component
functions of an {\it effective metric field}.
This is not cosmetics. Indeed, in some very special systems and approximations
the hydrodynamic equations governing the dynamic of the effective
metrics, when recast in geometric terms, can also be seen to
reproduce known geometrodynamic theories, at least in part,
ranging from Newtonian gravity to (almost) GR.  This first type of results is so far limited to special systems,
peculiar approximations, and ultimately not fully satisfactory, as it has not been possible yet to
reproduce, say, the Einstein-Hilbert dynamics
in any, however idealized, condensed matter system.
On the contrary, it is a very general result that the
effective dynamics of perturbations around the same background configurations turns out to be
given by matter field theories in curved spacetimes, whose
geometry is indeed the one identified by the effective metrics
obtained from the collective background parameters of the fluid. Moreover, the effective theories have often relativistic (if approximate) symmetries, even if the fundamental theory may be invariant only under galilean
symmetries. In this sense, we can speak of matter and Lorentz symmetry as emerging.
It is this type of results that we managed to reproduce, so far, in a GFT context, including the results of this paper.

The general scheme of what goes on in all these condensed matter systems, concerning the emergence of effective matter field theories, is well captured by the following \lq\lq meta-model\rq\rq (see \cite{analog-review} for details and for the exact assumptions entering the model).

One considers (for simplicity) a scalar field $\phi(\x)$, encoding the kinematical variables of the system (fluid) under consideration (e.g. velocity and density of a condensate in the hydrodynamic approximation) on  a classical continuum spacetime $\mm$ of
dimension $d$, with a dynamics governed by the lagrangian  $\lll(\phi,\partial_\mu
\phi,h_\mn)$ function of $\phi$ and its derivatives
$\partial_\mu \phi$ with respect to coordinates on $\mm$. Note
that if we introduce derivatives, we need to introduce tensors
such as a metric $h_{\mu\nu}$  to contract indices. A priori we do
not make any assumption on the signature of this metric tensor.

We introduce the solution $\phi_0(\x)$ of the equations of motion
induced by $\lll$, and then perturbations around
$\phi_0(\x)$.
\begin{equation}\label{perturbation}
\phi(\x)=\phi_0(\x)+\phi_1(\x).
\end{equation}
Expanding the Lagrangian around the classical solution chosen, we obtain:
\begin{eqnarray}\label{expansion}
\lll(\phi,\partial_\mu \phi,h_\mn) &= &\lll(\phi_0,\partial_\mu \phi_0,h_\mn)+  \left.\frac{\partial\lll}{\partial (\partial_\mu\phi)}\right|_{\phi_0}\partial_\mu\phi_1 +  \left.\frac{\partial\lll}{\partial \phi}\right|_{\phi_0}\phi_1  \label{linear-term}\\
&+& \frac{1}{2} \left( \left.\frac{\partial^2\lll}{\partial
(\partial_\mu\phi)\partial
(\partial_\mu\phi)}\right|_{\phi_0}\partial_\mu\phi_1
\partial_\mu\phi_1+2 \left.\frac{\partial^2\lll}{\partial
(\partial_\mu\phi)\partial \phi}\right|_{\phi_0}\partial_\mu\phi_1
\phi_1+ \left.\frac{\partial^2\lll}{\partial \phi\partial
\phi}\right|_{\phi_0}\phi_1 \phi_1\right) + ...
\end{eqnarray}
We can use a measure $\meas=\mmm dx^d= \mmm d\x$ on $\mm$ (where
$\mmm$ depends on the metric $h_\mn$, of course), and define an action for
the fundamental field $\phi$, which in turns implies an action for
the perturbations \cite{analog-review}.
\begin{equation}\label{action2}
\ss[\phi]= \int_\mm \meas\lll= \ss[\phi_0, \mmm]+
\frac{\epsilon^2}{2}\int_\mm
\meas\left[\frac{\partial^2\lll}{\partial
(\partial_\mu\phi)\partial (\partial_\mu\phi)}\partial_\mu\phi_1
\partial_\mu\phi_1 + \left(\frac{\partial^2\lll}{\partial
\phi\partial
\phi}-\partial_\mu\left(\frac{\partial^2\lll}{\partial
(\partial_\mu\phi)\partial \phi} \right)\right)\phi_1
\phi_1\right].
\end{equation}
The usual approach in the \emph{emergent gravity} framework is to
identify the \emph{emergent} (inverse) \emph{metric} $g^{\mn}$  and
measure $\dx_g$  (depending both on  $\phi_0$) in \eqref{action2}
as
\begin{equation}\label{emergent-metric}
(dx^d\sqrt{|g|})g^{\mn}= \dx_g g^{\mn}\equiv \meas
\left.\frac{\partial^2\lll}{\partial (\partial_\mu\phi)\partial
(\partial_\nu\phi)}\right|_{\phi_0}= \mmm dx^d
\left.\frac{\partial^2\lll}{\partial (\partial_\mu\phi)\partial
(\partial_\nu\phi)}\right|_{\phi_0},
\end{equation}
where $g=det g_\mn$.
The equation of motion for the perturbations from the action
\eqref{action2} read
\begin{equation}\label{EOM-perturbations1}
\partial_\mu\left[\frac{\partial^2\lll}{\partial (\partial_\mu\phi)\partial (\partial_\mu\phi)}\partial_\mu\phi_1 \right]  - \left(\frac{\partial^2\lll}{\partial \phi\partial \phi}-\partial_\mu\left(\frac{\partial^2\lll}{\partial (\partial_\mu\phi)\partial \phi} \right)\right)\phi_1=0.
\end{equation}
They can be rewritten as
\begin{equation}\label{EOM-perturbations2}
\left( \square_{\phi_0} - V_{\phi_0}(\phi_1)\right)\phi_1=0,
\end{equation}
 with the operator
$\square_{\phi_0}= g^{\mu\nu}\partial_\mu\partial_\nu$, for the
effective (inverse) metric $\sqrt{-g}\,
g^{\mu\nu}=\frac{\partial^2
L}{\partial(\partial_\mu\phi)\partial(\partial_\nu\phi)}\mid_{\phi_0}$, and we have a potential $V(\phi_1)$ also depending on the background field $\phi_0$. Note that the
d'Alembertian operator is then a self-adjoint operator with
respect to the measure $\dx$. It is the possible to invert to obtain the effective metric and
from this the other tensors characterizing the effective spacetime
geometry. Notice that both the \lq\lq fundamental\rq\rq field and
the quasi-particle one live on a 4-dimensional spacetime of
trivial topology, although endowed with a different metric in general. We refer to the literature for further details and
applications of the above general result. The main point to notice
here is that one generically obtains an effective spacetime
geometry to which the quasi-particles couple, depending only in
its precise functional form on the fundamental Lagrangian $L$ and
on the classical solution $\phi_0$ chosen; they do not
couple to the initial (laboratory) flat background metric.

This is the type of mechanism we have reproduced in a Group
Field Theory context. Assuming that a given GFT model (Lagrangian) describes
the microscopic dynamics of a
{\it discrete quantum spacetime}, and that some solution of the
corresponding fundamental equations can be interpreted as
identifying a given quantum spacetime configuration, we have obtained an effective macroscopic {\it continuum} field theory for
matter fields from it, and shown that the effective matter
field theories that we obtain most easily from GFTs are quantum field
theories on non-commutative spaces of Lie algebra type.
Let us notice that, while the
correspondence between classical solutions of the fundamental
equations and effective geometries is a priori unexpected in
condensed matter systems, which are non-geometric in nature, so that they are referred to as {\it analog} gravity models, in the GFT case the situation is different. We have
here models which are non-geometric and far from usual
geometrodynamics in their formalism, but which at the same time
are expected to encode quantum geometric information and indeed to
determine, in particular in their classical solutions, a (quantum
and therefore classical) geometry for spacetime \cite{gftdaniele,gftlaurent}, also
at the continuum level, the issue being how they exactly do so. We
are, in other words, far beyond a pure analogy.

However, there are also a number of differences which
could be of interest, between the condensed matter context and the GFT one we have analyzed. Some are more of a technical nature.
The definition of the emerging metric \eqref{emergent-metric}
comes from derivatives of the scalar field, in particular from its
kinetic term. In the GFT we consider, the kinetic
term is trivial, in that it does not contain derivatives of the field. We are however dealing with a non-local
theory (contrary to the lagrangian $\lll$), because of the combinatorics of group convolutions in the interaction term, and the propagator $
\mathcal{K}(g)$ is then generated from this non-local potential.

Usually in analog models one works with a space-time which is topologically $\R^4$. This background is used to define the fundamental field
$\phi$, but also the perturbations $\phi_1$ do see the same topological structure.
What emerges and differs from the background structure is only the metric that should be added
on the vector space $\R^4$ to turn it into a metric space. In the GFT model, the fundamental
field is defined on a very large space $\SO(4,1)^{\times
4}\sim\R^{40}$ or $\AN_3^{\times4}\sim\R^{16}$ according to whether we
consider the group to be $\SO(4,1)$ or $\AN_3$, and both are non-commutative spaces. We then look at 2d perturbations defined
on the space $\SO(4,1)\sim\R^{10}$ or $\AN_3\sim\R^{4}$, which are again
non-commutative spaces of a similar type but of much lower dimension. This dimensional reduction is  natural in GFT models as well as from the point of view of point particles coupled to gravity at the spin foam (GFT perturbative expansion) level. However, we are not aware of a similar phenomenon in usual analog models.
As a consequence, the symmetries of the fundamental theory are clearly larger than
the "emerging" $\ka$-Poincar\'e symmetry. This latter arose
because of the dimension reduction and the specific choice of
classical solution. It is quite possible that different
solutions would then generate different symmetry groups.
For example by choosing a different vector $v_0$, we can
have access to different subgroups of $\SO(4,1)$. This feature is
similar to some situations in analog models, where choosing different solutions can
generate different metrics and therefore different space-time with
different symmetry groups. For example in BEC, one can generate
the Schwarzschild space-time, FRW space-time or Minkowski
space-time according to the chosen features of the background
condensate. What is not considered, again, is that the choice of solution could have also  the additional
dimension reduction effect we could see in GFT models. For example, if the perturbation was
defined on $\R^{10}\sim\SO(4,1)$, introducing the specific
solution encoded in $F$   reduces the dynamical field content to
fields living on $\R^4$.  Similarly,  our choice of a perturbation field living on a subspace of the spacetime in which the fundamental GFT field lives has again no analog in condensed matter systems.

At a more conceptual level, we point out a further, important difference between our procedure and the situation in condensed matter analog models. Quasi-particle dynamics on effective emergent metrics is obtained, in analog gravity models, from the {\it hydrodynamics} of the fundamental system, i.e. from what is already an effective theory with respect to the fundamental description of the microscopic dynamics of the system (e.g. the non-relativistic quantum field theory for the atoms of Helium in superfluid Helium-3).  In our GFT context, on the other hand, we have applied a similar procedure directly to the fundamental microscopic field theory. One practical reason for this is simply that no coherent effective description of GFT models has been developed, even though its necessity has been argued for \cite{daniele}; another is that we have made crucial use of the fundamental definition of GFTs as field theories on group manifolds (suitably reinterpreted as momentum space) in order to obtain our effective matter fields living on non-commutative spaces of Lie algebra type. In fact, in the end this is why we have obtained {\it non-commutative} field theories for the emergent matter, as opposed to ordinary field theories, and one could speculate that this feature would be lost if the procedure was applied in some (to be sought for) effective GFT dynamics corresponding to some ordinary, continuum, classical geometrodynamics, thus such that it has lost memory of most of the {\it quantum} properties of spacetime at the Planck scale.

\section*{Conclusion}
We have derived
a scalar field theory of the deformed special relativity type, with a $\ka$-deformed Poincar\'e symmetry, from the $\SO(4,1)$ group field
theory defining the transition amplitudes for topological BF-theory in 4 spacetime dimensions. This was done directly at the GFT level, thus bypassing the corresponding spin foam formulation, in such a way that matter fields emerge from the fundamental  model as perturbations around a specific phase of it, corresponding to a solution of the fundamental equations of motion, and the non-commutative field theory governs their effective dynamics.
Not only this is the first example of a derivation of a DSR model for matter from a more fundamental quantum gravity model, and one further link between non-commutative geometry and quantum gravity formulated in terms of spin foam/loop quantum gravity ideas, but it is of great interest from the point of view of quantum gravity phenomenology, as we have pointed out in the introduction. It represents, in fact, another possible way of bridging the gap between quantum gravity at Planck scale and effective (and testable) physics at low energies.

Obviously, the are many questions left unanswered in this work. Some concern purely technical details of our procedure. We have mentioned them in the bulk of this paper, so we do not repeat them.
We mention here briefly a few more general ones of these open issues.

The first concern the role of the $\SO(3,1)$ degrees of freedom in the group field theory we started from, as well as in the one we have obtained as describing the dynamics of matter. From the GFT point of view it is utterly unclear why $\AN^c_3$ should be the relevant group for the perturbations as opposed to some other subgroup of $\SO(4,1)$. One can pose this same question in terms of the classical solution we have perturbed around. What is the physical meaning of the solution we have chosen? This is unclear at present, contrary to the 3d case, where the solutions used an be related to flat geometries. As mentioned, we expect it to be related to de Sitter space, but more work is needed to understand the details of the correspondence.
Related to this, it would be interesting to investigate the role of the cosmological constant in this GFT context. To start with, it seems that here the presence of a cosmological constant is encoded only in the group manifold used in the starting GFT, i.e. $\SO(4,1)$, but we have little control of how this is done. Second, we have motivated the choice of starting with this GFT model also by analogy with the McDowell-Mansouri (and related) formulation of General
Relativity as a $\SO(4,1)$-gauge theory, but this works only for a strictly positive cosmological constant. It is then natural to ask what happens if we start from $\SO(3,2)$ in place of $\SO(4,1)$ in the original model and then carry out the same procedure for extracting an effective matter field theory.
Further investigations are needed to establish a better link between our initial GFT model, classical solutions and effective field theory on the one hand, and a spin foam formulation of the Freidel-Starodubstev classical gravity theory \cite{mm} and the particle observable insertions  {\it \`a la} Kowalski-Glikman-Starodubtsev \cite{artem} on the other, which represent another path to deriving an effective deformed special relativity from spin foam models.
Last, we have obtained a scalar field theory for matter, and thus we should now look for extensions of our procedure and result that could give instead matter fields with non-zero spin e.g. Dirac fermions or vector fields. Moreover, we have provided an example of the emergence of spacetime (deformed) isometries from GFT, but it is natural to wonder if also gauge symmetries and thus gauge fields can be seen as emerging from some fundamental (GFT) quantum gravity model. Higher spins have already encoded in 3d GFT in \cite{iojimmy}, but never in 4d and in the usual sense of {\it coupling} matter degrees of freedom to quantum gravity ones, instead of having the first emerge from the second, as in the present work. Therefore, this is an area of research that is still wide open to be explored.
We leave all these questions for future work.

%
%
\section*{Acknowledgements}
We thank L. Freidel, J. Kowalski-Glikman, S. Liberati, B. Schroers, L. Sindoni for discussions and comments.
D. O. gratefully acknowledges financial support from the Alexander Von Humboldt Foundation
through a Sofja Kovalevskaja Prize.


\end{document}